# Heterogeneous Condensation On Nanoparticle

Sergey P. Fisenko[1], Manabu Shimada[2] and Kikuo Okuyama[2]

[1]*A.V. Luikov Heat& Mass Transfer Institute, National Academy of Sciences,*

*15, P. Brovka St., Minsk, 220072, Belarus*

E-mail:fsp@hmti.ac.by

[2] *Department of Chemical Engineering, Graduate School of Engineering,*

*Hiroshima University, Higashi Hiroshima, 739-8527 Japan*

**Abstract** Vapor condensation on nanoparticle with radius smaller than the Kelvin radius is considered as fluctuation or as the heterogeneous nucleation. The expression for steady-state heterogeneous nucleation rate is obtained. Nucleation on negatively charged nanoparticles is discussed. The report was made at 17[th] International Nucleation and Atmospheric Conferences, Galway, Ireland, 2007.

## INTRODUCTION

From thermodynamic point of view a vapor condensation starts on the spherical nanoparticle if the radius $R$ of the nanoparticle is greater then the Kelvin radius, $R_c$. The latter radius is determined by means of the expression [1]:

$$R_c = 2\sigma v_a / kT \ln(S), \tag{1}$$

where $\sigma$ is the surface tension of the condensed liquid, $v_a$ is the volume per molecule in this liquid, $S$ is the vapor supersaturation, $k$ is the Boltzmann's constant, $T$ is the temperature of the system. If nanoparticle radius is smaller the Kelvin radius the condensation is prohibited from thermodynamic point of view. Nevertheless due to fluctuations the vapor condensation is possible, we call this process as heterogeneous nucleation. Basic parameters of heterogeneous nucleation are the main subject of our report. It is worthy to emphasize that insight into heterogeneous nucleation is practically important for evaluation of the performance of particle size magnifiers (PSM) [2,3].

## FREE ENERGY OF WETTING FILM FORMATION

The free energy of formation of wetting film $\Delta\Phi(R,g)$, which has $g$ molecules, on the surface of the spherical nanoparticle can be written as:

$$\Delta\Phi(R, g) = -gkT \ln(S) + 4\pi \left[ R^2(\sigma_{sl} - \sigma_{sv}) + R_1^2 \sigma \right] \tag{2}$$

where $\sigma_{sl}$ and $\sigma_{sv}$ are, correspondingly, the surface tension between the solid nanoparticle and the liquid, and the surface tension between the solid nanoparticle and vapor, $R_1$ is the outer radius of wetting film, we will call it the shell radius. In particular for system of gold - water the contact angle is equal to zero, and for calculation $\sigma_{sv}$ we have the condition:

$$\sigma_{sv} = \sigma + \sigma_{sl}.$$

For core gold nanoparticle with radius $R$=2.5nm and water vapor supersaturation $S$=1.4, the free energy of wetting film formation versus the radius of the shell is displayed in Fig. 1. The existence of thermodynamic barrier of wetting film formation is obvious. The height of this barrier (~50$kT$) is typical for homogeneous vapor nucleation. The position of the maximum of the thermodynamic barrier exactly corresponds to the Kelvin radius $R_c$. Qualitatively it is clear from data in Fig.1 that overcoming of thermodynamic barrier is important if the nanoparticle radius is slightly smaller than $R_c$. Analytic results are presented below.

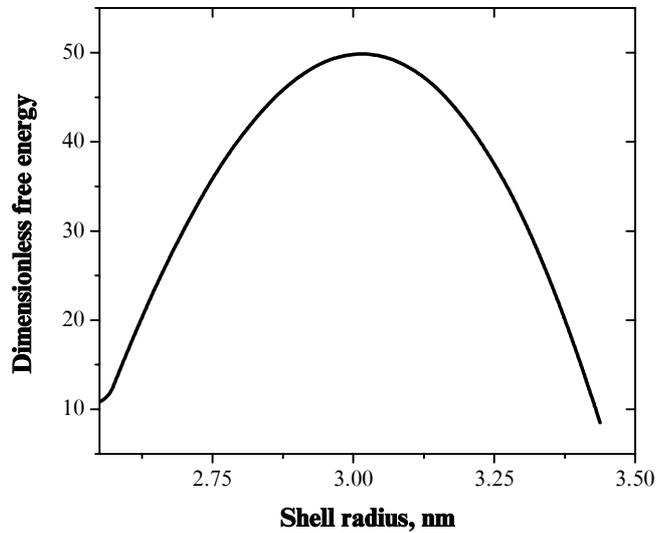

**Figure 1** Dimensionless free energy of the liquid shell formation $\Delta\Phi(2.5nm,g)/kT$ versus shell radius $R_1$

Following [5], the free energy of negatively charged nanoparticles can be presented as following:

$$\Delta\Phi_e(R,g,q) = \Delta\Phi(R,g,) + \frac{q^2}{8\pi\varepsilon_0}\left[\frac{1}{R_1} - \frac{1}{R}\right]$$

where $q$ is the charge of nanoparticle, $\varepsilon_0$ is the vacuum dielectric permittivity. For single charged nanoparticles with radius 2nm, the free energies differ only on several kT units. Electrostatic contribution to the free energy is very small if nanoparticles radius is larger than 4nm.

## KINETICS OF HETEROGENEOUS NUCLEATION

Let us introduce the ensemble of heterogeneous droplet with core nanoparticles of the same size and corresponding the distribution function $f(g)$ of the number molecules $g$ in the liquid shell over the nanoparticle. We assume below that $g$ is the continuous variable.

The kinetic equation for distribution function is the continuity equation, which includes the Brownian diffusion in the space of the number of molecules is the liquid shell. It can be shown that the kinetic equation for heterogeneous nucleation is formally similar to the kinetic equation for homogeneous nucleation [4]:

$$\partial_g\left[f(g)L_{11}\partial_g\left(ln\,f(g) + \beta\Delta\Phi(R,g)\right)\right] = -\partial_g[J_h] \qquad (3)$$

where $L_{11}$ is the free molecular flux of vapor molecules on the surface of the critical heterogeneous droplet, which radius coincide with Kelvin's radius; $J_h$ is the rate of heterogeneous nucleation. The physical meaning of $J_h$ is the flux of heterogeneous droplets overcoming thermodynamic barrier of the heterogeneous condensation.

Boundary conditions to the kinetic equation (3) are:

for small $g$ we have equilibrium with adsorbed molecules:

$$f(g) = C\,exp[-\beta\Delta\Phi(R,g)];  \qquad (4)$$

for large $g$ we have the standard condition:

$$f(g) = 0,$$

where $C$ is the normalization constant. The normalization constant $C$ is related with the number density of nanoparticles $N_p(g)$ with $g$ molecules in liquid core by the relationship:

$$N_p(g) = C\,exp(-\beta\Delta\Phi(R,g))\,dg \qquad (5)$$

After approximate integration of the both sides of (5) over $g$ we have the expression:

$$C \simeq N_t\,exp(\beta\Delta\Phi(R,0)),$$

where $N_t$ is the total number of nanoparticles in the ensemble.

The heterogeneous nucleation rate $J_h$ is the first integral of the kinetic equation (3), and after calculations we have the formula:

$$J_h = N_t L_{11}\,exp\left[-\beta(\Delta\Phi(R,g^*) - \Delta\Phi(R,0))\right]\sqrt{\alpha/\pi} \qquad (6)$$

where $g^*$ is the number of molecule at the critical shell, and

$$\alpha = 0.5 \left| \partial_{gg}^2 \Delta \Phi(R, g^*) \right|.$$

To remind that *g\** depends on radius of the nanoparticle, temperature, supersaturation, and thermophysical properties of the liquid. In particular for conditions shown in Fig.1 heterogeneous nucleation gives substantial effect if the nanoparticle radius is larger $0.95 R_c$.

## CONCLUSIONS

It was shown that vapor condensation can start on the surface of the nanoparticle even if its radius smaller than the Kelvin radius. This process is stochastic process, and the probability to overcome thermodynamic barrier is smaller than one. The free energy of wetting film formation is given by the expression (2). The flux of heterogeneous droplets, which overcome the thermodynamic barrier, is the rate of the heterogeneous nucleation. For calculation of heterogeneous nucleation rate the formula (6) was obtained. If the radius of core increases the heterogeneous nucleation rate drastically increases because free energy of wetting film formation drops.

The heterogeneous nucleation on charged nanoparticles is also discussed in the frame developed kinetic description.